\documentclass[aps,prl,amsmath,amssymb,twocolumn,reprint]{revtex4}
\usepackage{epsfig,amssymb}
\usepackage{graphicx}
\usepackage{pict2e}
\usepackage{dcolumn}
\usepackage{bm}
\usepackage{dcolumn}
\usepackage{multirow}
\begin{document}
\title{Satellite structures in the spectral functions of semiconductor
  quantum well two-dimensional electron gases: a GW
  plus cumulant study}

\author{Johannes~Lischner, Derek Vigil-Fowler, and Steven~G.~Louie}

\affiliation{Department of Physics, University of California,
  Berkeley, California 94720, USA, and Materials Sciences Division,
  Lawrence Berkeley National Laboratory, Berkeley 94720, USA.}

\begin{abstract}
  We present theoretical calculations for the spectral functions and
  single-particle densities of states of the two-dimensional electron
  gas in semiconductor quantum wells at different electron densities
  using the GW+cumulant method.  We compare our results to GW only
  calculations and find significant differences in the description of
  the satellites between the two theories: while GW theory predicts
  the existence of a plasmaron excitation, no such excitation is found
  in GW+cumulant theory. We compare our results to experimental
  tunneling spectra from semiconductor quantum wells and find good
  agreement for the satellite properties.
\end{abstract}

\maketitle

\emph{Introduction}.---The two-dimensional electron gas (2DEG) is of
fundamental importance to condensed matter physics and has been
studied extensively over the past decades. Experimentally, this system
can, for example, be realized at a semiconductor heterojunction. Many
new phenomena, including the integer \cite{VonKlitzing} and fractional
quantum Hall effects \cite{Stormer}, were discovered in this
system. Recently, there has been much interest in 2DEGs at the
interface of oxide materials\cite{Ohtomo,Ahn}. Also the theoretical
study of the 2DEG has a long history. Due to its conceptual simplicity,
this system has served as a test bed for many-body theories of
interacting electrons, such as Green's function theories
\cite{DasSarma,Ando} or Quantum Monte Carlo
studies\cite{Ceperley1,Needs}.

Recently, Dial and coworkers used time-domain capacitance spectroscopy
\cite{DialNature1,DialNature2} to measure the single-particle density
of states (DOS) of a 2DEG with high accuracy \cite{Dial}. As expected
from previous Green's function calculations based on the GW
approximation to the electron self energy \cite{DasSarma}, the
experimental DOS of Dial et al. exhibits two features: at low binding
energies, a structure due to quasiparticle excitations is observed
and, at higher binding energies, a second feature attributed to
plasmon satellites is observed as well. However, the onset of the
satellite structure in the experiment disagrees with theoretical
results from GW theory: GW significantly overestimates the separation
between the edges of the quasiparticle and satellite structures.

The failure of GW theory to describe satellites accurately has been
found in other systems as well. For the photoemission spectrum of a
core state, Langreth \cite{Langreth} demonstrated that the GW theory
severely overestimates the quasiparticle-satellite separation and also
results in only a \emph{single} satellite instead of a infinite series
of satellites. Recently, \emph{ab initio} studies
\cite{Guzzo,Lischner,LischnerSilicon} reported a similar
overestimation of the satellite binding energy for the valence band
photoemission spectrum of silicon, and it is found\cite{Lischner} that
GW theory does not describe the satellites in angle-resolved
photoemission studies of doped graphene \cite{Rotenberg2} accurately.

These studies \cite{Langreth,Guzzo,Lischner} moreover showed that a GW
plus cumulant (GW+C) theory \cite{Hedin} which includes significant
vertex corrections beyond GW cures the deficiencies of the GW theory
for the description of plasmon satellites and gives good agreement
with experiments. In particular, while GW predicts a new well-defined
excitation in the spectral function, the \emph{plasmaron}
\cite{Lundqvist,Rotenberg2}, resulting from the electron-plasmon
coupling, our studies \cite{Lischner} showed for both silicon and
doped graphene that no such plasmaron solution exists within the GW+C
theory, indicating the spurious nature of the plasmaron.

In this article, we present the first application of the GW+C theory
to the 2DEG at a semiconductor heterojunction. Also, for this system,
previous calculations based on the GW approximation alone found a
plasmaron solution \cite{Dial,DasSarma}. We present spectral functions
and electronic DOSs for different electron densities and compare our
results to GW calculations and also to the experimental findings of
Dial and coworkers \cite{Dial}. We do not find a plasmaron solution in
the GW+C theory.

\emph{Methods}.---The current in a tunneling experiment where a
voltage difference $V$ is applied across two systems with DOS
$g_l(\epsilon)$ and $g_r(\epsilon)$ within the Bardeen formalism is given
by\cite{Kaxiras}
\begin{equation}
I(V) = \int_{\epsilon_F}^{\epsilon_F + eV} d\epsilon g_l(\epsilon)g_r(\epsilon+eV) |M(\epsilon)|^2,
\end{equation}
where $M(\epsilon)$ is the tunneling matrix element and $\epsilon_F$
and $e$ denote the Fermi energy and the electron charge,
respectively. In time-domain capacitance spectroscopy, one system is a
three-dimensional electrode and the other a 2DEG. If both the DOS of
the three-dimensional electrode and the tunneling matrix element are
slowly varying functions of $\epsilon$, the derivative of the current
with respect to $V$ is proportional to the DOS of the 2DEG.

The many-body DOS per unit area of a two-dimensional paramagnetic
system with a single band, such as an ideal 2DEG, is given by
$g(\epsilon)= 2\int d^2k/(2\pi)^2 A_{\bm{k}}(\epsilon)$ with
$A_{\bm{k}}(\epsilon)$ being the many-body spectral function which is
related to the interacting one-particle Green's function via
$A_{\bm{k}}(\epsilon)=1/\pi |\text{Im} G_{\bm{k}}(\epsilon)|$.

Usually, $G_{\bm{k}}(\epsilon)$ is obtained by solving Dyson's equation
$G^{-1}_{\bm{k}}(\epsilon) =
G^{-1}_{0,\bm{k}}(\epsilon)-\Sigma_{\bm{k}}(\epsilon)+V^{xc}_{\bm{k}}$
with $G_{0,\bm{k}}(\epsilon)$ and $V^{xc}_{\bm{k}}$ denoting a
mean-field Green's function and mean-field exchange-correlation
potential, respectively, and $\Sigma_{\bm{k}}(\epsilon)$ is the self
energy for the state $\bm{k}$.

While describing quasiparticle properties in many materials with high
accuracy \cite{LouieHybertsen}, the GW approximation is less reliable
for satellite properties \cite{Langreth,Lischner,HedinAryasetiawan,Guzzo}: for
the spectral function of a core electron interacting with plasmons, GW
predicts a single satellite instead of a satellite series with
decreasing spectral weight and also greatly overestimates the binding
energy of the satellite structures. The cumulant expansion
\cite{Langreth,Guzzo,Lischner,HedinAlmbladh,Hedin,HedinAryasetiawan}
of $G_{\bm{k}}(\epsilon)$ cures these deficiencies by including
significant vertex corrections beyond GW: it provides the \emph{exact}
solution for a core electron interacting with plasmons
\cite{Langreth}. In the cumulant approach, the Green's function for a
hole is expressed as
\begin{align}
  G_{\bm{k}}(t) = i \Theta(-t) e^{-i\epsilon_{\bm{k}}t/\hbar + C_{\bm{k}}(t)},
  \label{Gcumulant}
\end{align}
where $\epsilon_{\bm{k}}$ denotes the mean-field orbital energy and
$C_{\bm{k}}(t)$ denotes the cumulant. This expression for the Green's
function is obtained after the first iteration of the self-consistent
solution of its equation of motion assuming a simple quasiparticle
form for the starting guess \cite{HedinAlmbladh}.  

The cumulant can be separated into a quasiparticle part
$C^{qp}_{\bm{k}}(t)$ and a satellite part $C^{sat}_{\bm{k}}(t)$
given formally in terms of the self-energy by (for $t<0$)
\begin{align}
  C^{qp}_{\bm{k}}(t) &= -it \Sigma_{\bm{k}}(E_{\bm{k}})/\hbar + 
  \frac{\partial \Sigma^h_{\bm{k}}(E_{\bm{k}})}{\partial \epsilon} \label{Cqp}\\
  C^{sat}_{\bm{k}}(t) &= \frac{1}{\pi} \int_{-\infty}^{\mu} d\epsilon 
  \frac{\text{Im}\Sigma_{\bm{k}}(\epsilon)}{(E_{\bm{k}}-\epsilon-i\eta)^2}
  e^{i(E_{\bm{k}}-\epsilon)t/\hbar},
  \label{Csat}
\end{align}
where $\mu$ denotes the chemical potential, $\eta$ is a positive
infinitesimal,
$E_{\bm{k}}=\epsilon_{\bm{k}}+\Sigma_{\bm{k}}(E_{\bm{k}}) - V^{xc}_{\bm{k}}$
is the quasiparticle energy and $\Sigma^h_{\bm{k}}(\epsilon)$ is
defined through the relation
\begin{align}
\Sigma^h_{\bm{k}}(\epsilon)=\frac{1}{\pi}
\int_{-\infty}^{\mu} d\epsilon'
\frac{ \text{Im}\Sigma_{\bm{k}}(\epsilon')}{\epsilon'-\epsilon-i\eta}.
\label{Sigmah}
\end{align}

For a given level of approximation for $\Sigma$, the cumulant theory
yields an improved Green's function through
Eqs.~(\ref{Gcumulant}-\ref{Sigmah}). In the present study, $\Sigma$ is
obtained from GW theory \cite{LouieHybertsen,HedinBook} which is known to
describe quasiparticle properties in many materials accurately thus
providing a good starting point for the cumulant theory.

\begin{figure}
  \includegraphics[width=8.cm]{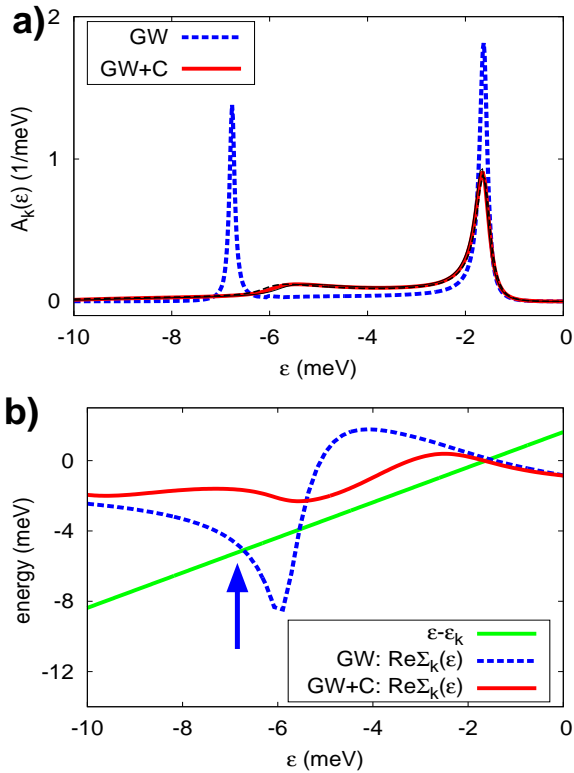} 
  \caption{(a): Spectral functions at $k=0$ for $n=5\times
    10^{10}$/cm$^2$ from GW and GW+C theories. Also shown are the GW+C
    spectral functions obtained by increasing (solid black line) and
    decreasing (dashed black line) the environment dielectric
    constant, the electrode distance and the well width by 10
    percent. (b): Real parts of the self energies from GW and GW+C
    theories. The arrow indicates the plasmaron solution in GW
    theory. All energies are measured with respect to the Fermi
    energy.}
  \label{fig:Ak0}
\end{figure}

\emph{Computational details}.---We use Hartree theory as the starting
mean-field theory, i.e. $V^{xc}_{\bm{k}}=0$. The bare Coulomb
interaction in an ideal 2DEG is $v(\bm{q})=2\pi e^2/|\bm{q}|$. We
include the finite width ($w=230~\AA$\cite{Dial}) of the electron gas
of the experimental sample by multiplying $v(\bm{q})$ with a form
factor \cite{DasSarma}. To account for the screening by the
surrounding dielectric environment we divide $v(\bm{q})$ by
$\epsilon_\infty=10.9$ corresponding to the GaAs/Al$_x$Ga$_{1-x}$As
heterostructure of Dial et al. \cite{Dial}. The metallic screening of
the distant electrode (being at a distance $d=245~\AA$ from the center
of the quantum well \cite{Dial}) is included by means of an image
charge model \cite{NeatonICM}. We have investigated the sensitivity of the GW+C
spectral functions on the environment screening, the electrode
distance and the well width by varying the parameters by 10 percent
and found the dependence to be quite weak [see Fig.~\ref{fig:Ak0}(a)].
Screening processes within the 2DEG are described via its analytically
known frequency-dependent polarizability function in the random-phase
approximation\cite{Ando}. We first calculate the imaginary part of the
self energy and then carry out a Kramers-Kronig transformation to
obtain the real part.

To improve the mean-field Green's function from Hartree theory, all
mean-field energies are shifted by a constant $\Delta$ which is
determined by requiring that $\hbar^2 k^2_F/(2m^*)+\Delta$ ($m^*=0.067 m$
denoting the effective electron mass \cite{Dial} with $m$ being the bare
electron mass and $k_F$ the Fermi wave vector) equals the quasiparticle
energy at $k_F$ obtained by solving Dyson's equation.  Finally, we
used the self energy from the GW calculation to evaluate the spectral
function in GW+C theory. We note that all results discussed below
include the effects of finite quantum well widths and dielectric
screening of the environment.

\emph{Results}.---Figure~\ref{fig:Ak0}(a) shows the spectral function
for a 2DEG with $n=5\times 10^{10}$/cm$^2$ at $k=0$, i.e., at the
bottom of the band, from GW and GW+C theories. Both spectral functions
exhibit two peaks: one quasiparticle peak at a lower binding energy
and a satellite peak at a higher binding energy. While the location of
the quasiparticle peak is the same in both theories, the location and
shape of the satellite peak is very different: GW theory gives a very
sharp satellite peak at an energy of $\sim -6.8$~meV, while GW+C
results in a much broader satellite structure with a weak peak at
$\sim -5.5$~meV.

Unlike the three-dimensional electron gas where the plasmons are
high-energy excitations with an energy scale of
$\omega_{pl}=\sqrt{4\pi n e^2/m}$ with $n$ denoting the electron
volume density, the plasmons in the 2DEG have a dispersion of
$\omega_{pl}(q) \approx \alpha(n) \sqrt{q}$ (in the long wavelength
limit without metallic screening from the distant electrode) with
$\alpha(n)=\sqrt{2\pi n e^2/(\epsilon_\infty m^*)}$. To understand the
separation between the satellite and the quasiparticle peaks in
Fig.~\ref{fig:Ak0}(a), we expand the exponential factor in
Eq.~\eqref{Gcumulant} into a power series in $C^{sat}_{\bm{k}}$, the
satellite contribution to the cumulant function. The resulting
spectral function is given by a quasiparticle contribution and an
infinite sum of satellites,
i.e. $A_{\bm{k}}(\epsilon)=A^{qp}_{\bm{k}}(\epsilon) +
\sum^{\infty}_{n=1} A^{sat,n}_{\bm{k}}(\epsilon)$ with
$A^{qp}_{\bm{k}}(\epsilon)$ being a Lorentzian centered at
$E_{\bm{k}}$ and $A^{sat,1}_{\bm{k}}(\epsilon)=\int d\epsilon'
C^{sat}_{\bm{k}}(\epsilon')A^{qp}_{\bm{k}}(\epsilon-\epsilon') \approx
Z_{\bm{k}} C^{sat}_{\bm{k}}(\epsilon-E_{\bm{k}})$ with $Z_{\bm{k}}$
denoting the renormalization factor. According to Eq.~\eqref{Csat},
$C^{sat}_{\bm{k}}(\epsilon-E_{\bm{k}})$ is proportional to the
imaginary part of the self energy which has a van Hove singularity
when the group velocity of the holes equals the plasmon group
velocity. For $k=0$, this argument predicts a separation of $\Delta
E=[\hbar^2 m^*\alpha(n)^4/2]^{1/3} \approx 4.6~$meV between the
quasiparticle peak and the satellite. This is somewhat larger, but
agrees reasonably well with the separation $\Delta E=3.8~$meV found in
Fig.~\ref{fig:Ak0}(a) from the full GW+C theory.

The discrepancy between GW and GW+C theory for the shape and location
of the satellite peak can be traced back to a spurious pole of the
Green's function in GW theory. Figure~\ref{fig:Ak0}(b) shows the
graphical solution of Dyson's equation for GW and GW+C theory. For
GW+C, we computed the vertex-corrected self energy via
$\Sigma_{\bm{k}}(\epsilon)=\epsilon-\epsilon_{\bm{k}}-G^{-1}_{\bm{k}}(\epsilon)$
with $G_{\bm{k}}(\epsilon)$ obtained from Eq.~\eqref{Gcumulant}. The
sharp satellite peak in the GW spectral function results from an
additional solution of Dyson's equation. Such an additional solution
on the real frequency axis was first found in the three-dimensional
electron gas by Lundqvist who introduced the term \emph{plasmaron}
describing a hole resonantly bound to plasmons\cite{Lundqvist}. No
such solution is found in the GW+C theory indicating the disappearance
of the plasmaron excitation when higher-order electron-electron
interactions are included. The origin of the spurious plasmaron
solution is thus a failure of GW theory, which replaces all satellites
peaks by a \emph{single, effective} satellite at an artificially high
binding energy \cite{Langreth}. Physically, the satellites in the GW+C
theory correspond to many-body states consisting of weakly interacting
hole-plasmon pairs.

\begin{figure}
  \includegraphics[width=9.cm]{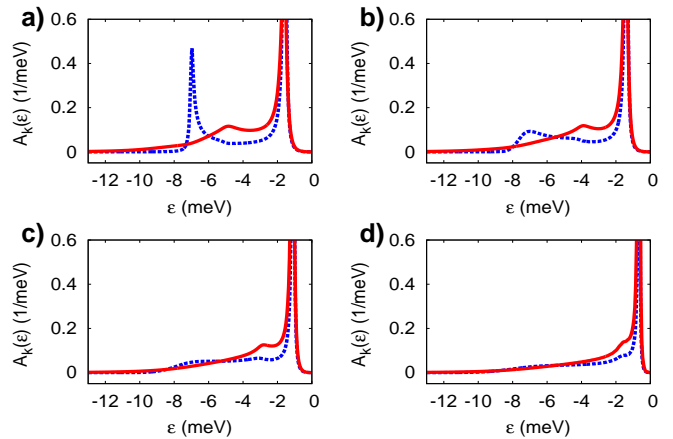} 
  \caption{Spectral functions at (a) $k=0.2~k_F$, (b) $k=0.4~k_F$, (c)
    $k=0.6~k_F$ and (d) $k=0.8~k_F$ for $n=5\times 10^{10}$/cm$^2$
    from GW+C (solid red curves) and GW (dashed blue curves)
    theories. All energies are measured with respect to the Fermi energy.}
  \label{fig:spectralfunctions}
\end{figure}

\begin{figure}
    \includegraphics[width=9.cm]{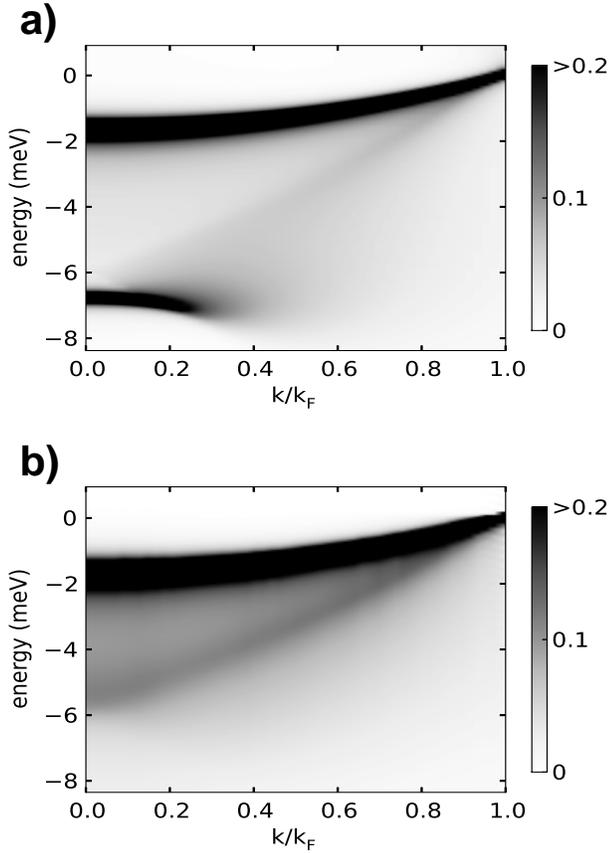}  
  \caption{Spectral functions $A_{\bm{k}}(\epsilon)$ (in units of 1/meV) of a
    two-dimensional electron gas with $n=5\times 10^{10}$/cm$^2$
    from (a) GW and (b) GW+C theories. All energies are measured with
    respect to the Fermi level.}
  \label{fig:contour}
\end{figure}

Figures~\ref{fig:spectralfunctions} and \ref{fig:contour} show
results for spectral functions at larger wave vectors:
Fig.~\ref{fig:spectralfunctions}(a) shows that, also at nonzero but
small wave vectors, GW theory gives a strong plasmaron
satellite. Fig.~\ref{fig:contour} shows that this GW plasmaron branch
disperses towards higher binding energies with increasing $k$ and then
disappears at $k\approx 0.25~k_F$. Interestingly, there is an
additional satellite branch in GW theory: it is much weaker than the
plasmaron branch and merges with the quasiparticle branch at $k_F$.

Figure~\ref{fig:contour}(b) shows that no plasmaron branch is found in
GW+C theory. We also find a satellite branch which merges with the
quasiparticles at $k_F$ and is somewhat stronger than the
corresponding weak feature in GW. Figure~\ref{fig:spectralfunctions}(d)
shows that near $k_F$ the spectral functions in GW and GW+C are more
similar than at small $k$.

\begin{figure}
  \includegraphics[width=9.cm]{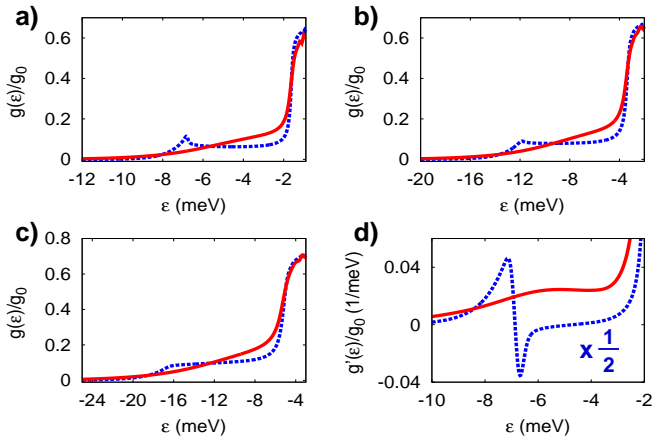} 
  \caption{DOS of a 2DEG with (a) $n=5\times 10^{10}$/cm$^2$, (b)
    $n=10 \times 10^{10}$/cm$^2$ and (c) $n=15\times 10^{10}$/cm$^2$
    from GW (dashed blue curves) and GW+C (solid red curves)
    theories. (d): $g'(\epsilon)=d g(\epsilon)/d\epsilon$ for
    $n=5\times 10^{10}$/cm$^2$. We have divided $dg/d\epsilon$ from GW
    by a factor of 2 to simplify the comparison with the GW+C
    result. $g_0=m^*/(\pi \hbar^2)$ denotes the value of the DOS
    neglecting electron-electron interactions. All energies are
    measured with respect to the Fermi energy. }
  \label{fig:dos}
\end{figure}

Figures~\ref{fig:dos}(a)-(c) show the resulting DOS from GW and GW+C
theories for different electron densities. The sharp increase at low
binding energies [for example, in Fig.~\ref{fig:dos}(a) at $\sim
-2$~meV] is due to the quasiparticle contribution to the DOS. Without
electron-electron interactions the DOS is a step function of height
$g_0=m^*/(\pi \hbar^2)$ with a sharp onset at $-\hbar^2
k^2_F/(2m^*)$. The feature at higher binding energies is caused by
plasmon satellites and is very different in GW and GW+C theories. GW
theory results in a plateau-like feature [between $\sim -7$ and $\sim
-2$~meV in Fig.~\ref{fig:dos}(a)] followed by a peak (arising from plasmaron
excitations) at -7 meV and a decaying tail. In the GW+C theory we do
not observe such a plateau in the DOS, but instead a tail which starts
at the edge of the quasiparticle feature and decays slowly with
increasing binding energy.

Direct comparisons of the computed DOS and experimental dI/dV curves
are difficult due to complicated matrix element effects
\cite{DialNature1}. To analyze the main features in the DOS, we follow
the procedure of Ref.\cite{Dial} and compute the derivative of the DOS
which --- assuming the electrode DOS and the matrix element are slowly
varying functions of $\epsilon$ --- is proportional to $d^2I/dV^2$
\cite{Dial}. Fig.~\ref{fig:dos}(d) shows that in GW theory the
plasmaron peak at the edge of the satellite structure causes a
characteristic feature in $dg/d\epsilon$: a sharp negative dip
followed by a positive peak as a function of increasing binding
energy. Instead, we find a broad peak in $dg/d\epsilon$ from GW+C
theory.

\begin{figure}
  \includegraphics[width=8.cm]{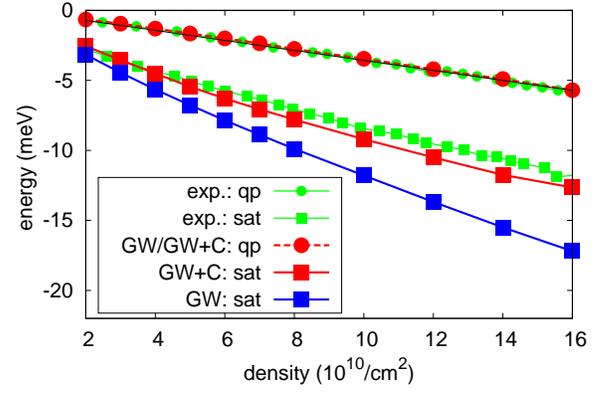} 
  \caption{Quasiparticle (qp) and satellite (sat) band edge positions
    from GW theory, GW+C theory and experiment \cite{Dial}. The
    dashed black line shows the quasiparticle band edge position from
    Hartree theory.}
  \label{fig:exp}
\end{figure}

Figure~\ref{fig:exp} shows the peak locations in $dg/d\epsilon$ for
different electron densities and compares the results of GW and GW+C
calculations with the experimental data of Dial and coworkers. For the
feature corresponding to the onset of the quasiparticle structure at
low binding energies, both GW and GW+C theories agree very well with
experiment. For the second feature at higher binding energy arising
from the satellite structure in the DOS, GW+C theory agrees much
better with experiment than GW theory. This indicates that the
plasmaron solutions in GW theory are indeed not physical.  We
attribute the remaining small difference between GW+C theory and
experiment to interactions with phonons. The longitudinal optical
phonon mode of gallium arsenide couples weakly to the electrons in the
2DEG. Das Sarma and coworkers have shown that this results in a small
decrease of the satellite band edge binding energy in GW theory and we
expect that phonons will have a similar effect in GW+C theory
\cite{DasSarma}.

In conclusion, we have carried out GW+C calculations for the
two-dimensional electron gas in GaAs quantum wells at various
densities and found good agreement with the experimental findings of
Dial and coworkers \cite{Dial}. Our results for the satellite
structures disagree qualitatively and quantitatively with calculations
based on the GW approximation only to the electron self energy. In
particular, we do not find a plasmaron excitation in GW+C theory.

J.L. acknowledges valuable discussions with Oliver Dial and Raymond
Ashoori. S. G. L. acknowledges support by a Simons Foundation
Fellowship in Theoretical Physics. This work was supported by NSF
Grant No. DMR10-1006184 (numerical simulations of the two-dimensional
electron gas) and by the Theory Program at the Lawrence Berkeley
National Laboratory funded by the Director, Office of Science, Office
of Basic Energy Sciences, Division of Materials Sciences and
Engineering Division, US Department of Energy under Contract
No. DE-AC02-05CH11231 (methods and software development of electron
correlation effects).  \bibliography{paper}
\end{document}